\documentclass[12pt,a4paper,english]{article}
\usepackage[T1]{fontenc}
\usepackage[latin2]{inputenc}
\usepackage[english]{babel}
\usepackage{amsmath}
\usepackage{amsfonts}
\usepackage{indentfirst}
\usepackage{epsfig}
\usepackage{graphicx}
\usepackage{color}
\begin{document}
\sloppy
\author{Szymon Niewieczerza{\l} and Marek Cieplak\\ \\ 
\textit{Institute of Physics Polish Academy of Science,}\\ 
\textit{Al. Lotnik\'ow \ 32/48, 02-668  Warsaw, Poland} \\
}
\title{
Stretching and twisting of the DNA duplexes in coarse grained dynamical models.
}

\maketitle
\bibliographystyle{unsrt}
\abstract{
Three coarse-grained  models of the double-stranded DNA are proposed and compared
in the context of mechanical manipulation such as twisting and various
schemes of stretching. The models differ in the number of effective
beads (between two and five) representing each nucleotide. They all show similar
behavior and, in particular, lead to a torque-force phase diagrams qualitatively
consistent with experiments and all-atom simulations.
}
\newpage
\section{Introduction}
Manipulation of large biomolecules by means of atomic force microscopy, 
optical tweezers, and other nanotechnological devices is playing an 
increasingly growing role in elucidating mechanisms of biologically 
relevant processes \cite{nat2003,topois,pnas1997}.
The dynamical data obtained through mechanical manipulation
usually requires theoretical 
interpretation that can be reached through numerical simulations.
This need is especially apparent when dealing with proteins 
(see, e.g., \cite{survey})
~because of their strongly inhomogeneous
network of interactions between the amino acids. The inhomogeneity
results, for instance, in a protein-dependent pattern of peaks  when the
force of resistance to pulling at constant speed is plotted against elongation
(see, e.g. \cite{Gaub,Clarke,Yang,Progress}).

All-atom simulations have contributed to understanding of the large 
conformational changes in proteins induced by the mechanical manipulation
(see e.g. \cite{Schulten,Paci,Amzelm}). However, such simulations are inherently
limited by the time scales and system sizes that can be studied. One way out
is to accelerate the modeled processes by orders of magnitude 
relative to their experimental
realizations which may, undesirably, distort the physics involved. 
Another way is to use coarse grained models. These models can also be applied to other
processes involving large conformational changes such as folding or 
thermal unfolding of biomolecules.
They may also find applications in studies of systems which are much 
larger in size.

In this paper, we focus on coarse grained models describing the double stranded
DNA (dsDNA). The simplest coarse graining scheme involves treating dsDNA as a  
polymer endowed with a local stiffness \cite{Siggia}. The resulting model seems 
to be appropriate in studies of stretching at high temperatures, when the 
entropic effects dominate, and in studies of phenomena related to fluid flow
and the hydrodynamic interactions \cite{Jendrejack}. The corresponding 
characteristics dimension
in this model can be measured by the Kuhn length or the hydrodynamic radius.
The typical values of these parameters are of order 106 and 77 nm \cite{Jendrejack} 
respectively which
encompasses more than 200 distances (of 0.4 nm ) between successive pairs of
nucleotides in the dsDNA. This level of coarse graining is thus too crude to be
useful when describing mechanical manipulation of the system where nanometric 
features are detectable. Here, we discuss coarse grained models with structures 
that are resolved at the level of a  single nucleotide.

There are a variety of possible ways to manipulate the DNA
duplex. The ones that we consider in this paper are schematically represented 
in Figure 1. Unzipping, whether at constant speed or at constant force, 
corresponds to scheme A in this Figure. The unzipping process involves
breaking of one hydrogen bond at a time resulting in a force, $F$, of
order 13-15 pN, that rapidly ondulates with the amplitude of order 1 pN
as the pulling distance, $d$, is increased \cite{unzip1:1997,unzip2:2002}.
The force pattern has also been found to depend on the pulling speed, $v_p$,
weakly \cite{unzip1:1997}, but would be expected to 
depend on temperature, $T$, more substantially. Even though the variations
in the $F-d$ patterns when pulling at constant speed take place on the
nucleotidic length scales, their interpretation in terms of the specifics
of the sequence is difficult because of the noise due to thermal fluctuations.
Recently, Baldazzi et. al. \cite{unzip4:2006t} has suggested, however,  that if 
instead the mechanical unzipping is performed at constant force then Bayesian 
methods of the corresponding sequence prediction should be nearly error-free.
Methods for extracting kinetic information from constant force experiments
have been discussed recently in the context of DNA unzipping in a nanopore
\cite{Dudko}.

Schemes B and C involve stretching at the opposite ends of the duplex.
The two schemes employ distinct mechanisms of resistance to
stretching. Mechanism B involves shear which is responsible for generation
of the strongest force clamps in biomolecules \cite{survey} 
and the 
maximum force obtained depends on the number of the bonds that are sheared 
simultaneously. Mechanism C, on the other hand, leads to localized  
unravelling and generates a size-independent force. When using 
micropipettes, as in the experiment by Cluzel et al. \cite{Cluzel}, one
probably combines schemes B and C. The resulting $F-d$ pattern has 
three stages: one starts off with a long period of a nearly constant force, 
after which there is a steady increase (in the 120 pN range) which finally
is followed by a sudden drop to zero. A similar pattern of behavior arises
in simulations involving anisotropic pressure \cite{luan}    
(these studies have been performed for a dsDNA with about 10 pairs of nucleotides).

Still another way of manipulating the dsDNA has been employed by
Oroszi et al. \cite{tors2:2006} and it involves applying a torque, $G$.
Wereszczynski and Andricioaei \cite{tors4:2007t} have generalized
it still further, in their all-atom simulations, by considering
a simultaneous application of a force and torque, as shown in scheme D
in Figure 1. They have predicted existence
of a rich phase diagram of possible structures on the $F-G$ plane.
There have been experimental studies involving torque produced
in an optical trap \cite{tors1:2003,tors2:2006,tors3:2007}.
Bryant et. al. \cite{tors1:2003} have found
that the torque needed to transform the B-DNA conformation into the
left-hand twisted L-DNA form is $G = -9.6$~pN~nm
(the negative sign indicates twisting against the native sense
of turn in the dsDNA),
while to transform it further
into the Pauling-like (P-DNA) form the required torque is $\;34$~pN~nm.
They have also observed that by pulling the dsDNA molecule with the
force of about 65~pN one induces a transition from the B-DNA into
the S-DNA form which is 60\% longer then the B-DNA.  

In this paper, we construct three variants of the nucleotide-based coarse 
grained dynamical models of the DNA duplex. The coarse graining method 
introduces several effective objects, referred here as beads, that represent 
a single nucleotide. The models differ primarily by the number of beads 
involved. We compare the workings of the three models for a 22-base-pair 
system, or shorter, and use them to elucidate the mechanisms 
of rupture in processes corresponding to schemes A through D.
One conclusion of our studies is that even though various dynamical details
differ between the models, all of them can be considered adequate and
ready to be applied to larger systems.
In particular, all of the coarse grained models studied lead to a
transitions of the usual right-hand-twisted B-dsDNA form to
the L-dsDNA form and to the P-DNA form on application of a  appropriate torque.

\section{An overview of the models used}

Physical properties of the DNA double helix are quite distinct \cite{Albert} 
compared to other biomolecules.
Its strong stiffness comes from the braided nature of its structure
combined with the presence of the base stacking interactions.
Furthermore, the phosphate groups in the
DNA backbone carry substantial electric charges. All of these
features are employed by the cell's machinery in the processes of
copying, transcribing and packaging  of the DNA. For example, 
helicases which unwind the double helix to provide single-stranded
templates for polymerases, have evolved as motors that are
capable of moving along the torsionally constrained DNA molecules.
Topoisomerases break and reconnect the DNA to relieve a torsional strain 
that accumulates ahead of the replication fork. Finally, the DNA-binding 
proteins get docked to the DNA by means of guidance mechanisms which
seem to be primarily electrostatic in nature.

Our models address the mechanical properties of the dsDNA and do
not aim at determining the electrostatic potential outside of the duplex.
The models are built in analogy to the
Go-like models of proteins, especially in the specific implementation proposed
in refs.\cite{biophysical,mc:prot2004,Kwiecinska,survey}. In the case of proteins,
the model represents the system by its C$^{\alpha}$ atoms which are tethered together
by harmonic interactions. The native contacts,
such as the hydrogen bonds, are described by the Lennard-Jones
potentials. The Langevin overdamped
thermostat with random forces mimics fluctuational effects of the solvent.
61 other variants of this basic model of a protein are discussed 
and compared in ref. \cite{models}.

The dsDNA has a simpler elastic structure than proteins
since a pair of nucleotides can bind only in two ways: either by forming
two (A-T) or three (G-C) hydrogen bonds. 
When trying to build a coarse grained model for a DNA
one is first inclined to assign a single bead to a nucleotide and to
locate it at the phosphorus (P) atom. This may be acceptable for a single
strand DNA, provided the local chain stiffness terms are included.
However, for the dsDNA this procedure would lead to a distance of 17 {\AA}
between the P atoms in a hydrogen bonded pair of the nucleotides.
Such a relatively large distance would introduce too much mechanical
instability in the model  but appears to be adequate to study 
conformational changes in dsDNA nanocircles and submicron-sized
plasmids with torsional stress \cite{Tozzini}.
A more detailed approach, denoted here as model I, 
involves representing the A and T nucleotides by four beads 
and the G and C nucleotides by five beads. One of the beads represents the
phosphate group, another the sugar group, and the remaining
beads participate in formation of either two or three hydrogen
bonds, depending on the specificity. 
The hydrogen bond interactions are represented
by the effective Lennard-Jones potentials and other bonds,
being structural in nature, 
are described by the harmonic potentials with large elastic constants.
The schematic construction of this model is shown in Figure 2.

More simplified approaches involve reduction in the number of beads 
representing each nucleotide. In a model denoted here as model II, we mimic 
the ribose-phosphate groups by one bead and the base by another bead as 
illustrated by the top of Figure 3. In this model, the distinction between
the A:T and G:C pairing interactions comes  not through introduction of 
separate contacts for each hydrogen bond but through adjustment of the 
amplitude of the effective base-base Lennard-Jones potential
by a factor of 2 or 3 respectively.

In between models I and II there is another model, denoted here as
model III, that has been introduced by Knott et al. \cite{pablo}
in the context of the salt induced melting.
Model III involves three beads as illustrated in the 
bottom part of Figure 3.
The beads represent the phosphate, sugar, and base groups correspondingly.
In model III, the backbone chain of one strand
is constructed by linking the sugar group
to the P atom on the same nucleotide and to the P atom on a preceding
nucleotide. Thus the backbone chain has the appearance of a zigzag line.
In contrast, in models I and II, the backbone chain is formed by
tethering the consecutive P atoms.

As a model system we consider the structure coded as 119D in the Protein
Data Bank which has been determined by Leonard and Hunter \cite{pdb:119D}.
It corresponds to the sequence
5'-D(CGTAGATCTACGTAGATCTACG)-3'. 
The ground state conformations of this system as captured by the three
models are shown in Figure \ref{models}.
Longer structures can  be obtained, for instance, by repeating this basic unit,
or by constructing a synthetic sequence-dependent structure that makes use 
of average geometrical parameters associated with a single base pair.

\section{Model I: the 4- or 5-bead description}

We start by introducing three different types of beads, p, h, and b
as illustrated in Figure~\ref{pbh2}.
The p-beads are meant to represent the backbone which is made of the
phosphate groups. The p-bead is placed at the position of the C4* atom
in the molecule of ribose. This placement achieves two goals.
First, it represents the DNA phosphate chain by the p-beads.  Second, it
locates the p-bead close to the base beads. The C4* atom
is the ribose ring atom that is closest to the phosphate group. 
The h-beads represent the 'head' atoms which may act either as
donors or as acceptors in the hydrogen bonds. In the C nucleotides,
the h-beads are located on the O2, N3, and N4 atoms of the bases.
In the G nucleotides, -- on the O6, N1, and N2 atoms. Finally, in the
A and T nucleotides -- on the N6, N1, and N3, O4 atoms respectively.
The h-beads are linked to their 'bases', i.e. to the supporting b-beads. 
In the native state, the b-beads are located half-way between the p-bead
and the center of mass of the h-beads at each nucleotide.
The overall scheme results in about  4- to 5-fold reduction in the
number of the degrees of freedoms compared to the all-atom approach.

In Table~\ref{tabela} we provide the Cartesian coordinates of the 
beads that define the model. Knowledge of these coordinates
should allow for a generation of a synthetic dsDNA based on
the sequence. The coordinates have been obtained by making averages
of the geometric parameters in the PDB structure corresponding to the 
basic sequence studied.

The p-beads are tethered into two separate chains and thus form two 
backbones. The tethering is accomplished through the elastic potential
\begin{equation}
V^{pp}_{i,i+1}= K_b\cdot(\vec{r}_{{p}_i} - \vec{r}_{p_{i+1}} - d_{p_{i}p_{i+1}})^2 \;\;,
\end{equation}
where index $i$ enumerates consecutive nucleotides and $d_{p_ip_{i+1}}$
are the distances between the consecutive p-beads in the native state. These distances
vary from  bead to bead. Their mean is equal to 5.8 {\AA}
and the standard deviation is close to 0.3{\AA}.
The mean geometrical parameters cited in this description can
be used in a general sequence-dependent construction of a 
synthetic dsDNA structure.
The elastic constant is taken to be equal to
$K_b=50\;\epsilon$\AA $^{-2}$, where $\epsilon$ is the energy scale 
corresponding to the internucleotidic hydrogen bonds, as defined below.

The steric constraints of the DNA sugar-phosphate backbone 
are represented by the following two potentials 
for the bond and dihedral angles of p-beads' backbone.
\begin{equation}
V^{B}= \sum_{i=\mathcal{N}_1}^{\mathcal{N}-2} K_{\theta}(\theta_i-\theta_{0i})^2 \;\;,
\end{equation}
\begin{equation}
V^{D}= \sum_{i=1}^{\mathcal{N}-3}[K^1_{\phi}(1+cos(\phi_i-\phi_{0i}))+K^3_{\phi}(1+cos3(\phi_i-\phi_{0i}))] \;\;.
\end{equation}
The bond angle $\theta_i$ is measured between the $p_i-p_{i+1}$ and 
$p_{i+1}-p_{i+2}$ bonds, and the dihedral angle $\phi_i$ is an angle between 
two planes: one of them is determined by  the $p_i-p_{i+1}$ and $p_{i+1}-p_{i+2}$ 
bonds, and the second one by the $p_{i+1}-p_{i+2}$ and $p_{i+2}-p_{i+3}$ bonds,
the subscript $0$ indicates the native values, $\mathcal{N}$ denotes 
the number of nucleotides in one chain.
We take $K_{\theta}=20\epsilon/(rad)^2$, $K^1_{\phi}=1.0\epsilon$, 
$K^3_{\phi}=0.5\epsilon$ in analogy to ref. \cite{Clementi}.

All of the interbead interactions within one nucleotide are taken
to be harmonic so that
the corresponding potentials read
\begin{equation}
V^{\nu \mu}_{i,i}= \sum_{\nu, \mu}  K_b\cdot(\vec{r}_{\mu_i} - \vec{r}_{\nu_i} - d_{\mu_i \nu_i})^2 \;\;,
\end{equation}
where the indices $\nu$ and $\mu$ label beads belonging to the $i$'th 
nucleotide. The equilibrium distances $d_{\mu \nu}$ take values as in the 
native structure and they range from 2.3 $\pm$ 0.1 {\AA} for the
neighboring h-beads (according to notation in Figure~\ref{pbh2},
pairs: $h_{1j}$-$h_{2j}$ and $h_{2j}$-$h_{3j}$)
to 3.6 $\pm$ 0.7 {\AA} between the b-bead and h-beads.

The h-beads on one chain are capable of making hydrogen-bond contacts with
the h-beads on the opposite chain. In the simplest version,
we follow the prescription used previously for proteins and describe these
contacts by the Lennard-Jones potential
\begin{equation}
V^{hh}_{i,j}= 4\epsilon [(\frac{\sigma_{h_ih_j}}{r_{h_ih_j}})^{12}
-(\frac{\sigma_{h_ih_j}}{r_{h_ih_j}})^{6}] \;\;,
\end{equation}
where $i$ and $j$ are the paired residues and
$r_{h_ih_j}=|\mathbf{r}_{h_i}-\mathbf{r}_{h_j}|$. 
The parameters 
$\sigma_{h_ih_j}$ are chosen so that each contact 
in the native conformation is stabilized in the minimum of the potential.
Essentially, $\sigma_{h_ih_j}=2^{-1/6}d_{h_ih_j}$. The value of 
$d_{h_ih_j}$, the distance between h-beads making bond, is equal to 2.66 $\pm$ 0.14~{\AA}.
For proteins, the choice of the form of the contact potential has turned out
to be of much less importance than the correct determination of 
the contact map \cite{models}.

The stacking interactions between consecutive b-beads in each chain are also
accounted for by the Lennard-Jones interactions:
\begin{equation}
V^{bb}_{i,i+1}= 4\epsilon [(\frac{\sigma_{b_ib_{i+1}}}{r_{b_ib_{i+1}}})^{12}
-(\frac{\sigma_{b_ib_{i+1}}}{r_{b_ib_{i+1}}})^{6}] \;\;.
\end{equation}
The distance between the stacking pairs of the b-beads 
is 4.43 $\pm$ 0.42 {\AA} in the native structure.   

All of the interactions discussed above arise in the native state. However,
distorted conformations may lead to new interactions. In the spirit of the
Go-like models, we describe these by purely repulsive potentials (the Lennard-Jones
potentials which are cut at the minimum and shifted). The hard sphere diameters  
of the h- b- and p-beads
are taken to be equal to 2.0 {\AA}, 3.4 {\AA},  and 6.0 {\AA} respectively.
The large effective size of the p-bead prevents the chains from crossing
and self-crossing.

All beads are endowed with the same mass, $m$, and the equation of motion
of each is described by the Langevin equation
\begin{equation}
m\ddot{\vec{r}} = -\gamma(\dot{\vec{r}}) + \vec {F}^c + \vec{ \Gamma} \;\;,
\end{equation}
which provides thermostating and mimics dynamical effects of the solvent.
Here $\vec{r}$ is the position of the bead, $\vec{F}^c$ is the net force 
on it due to potentials, $\gamma$ is the friction coefficient, and
$\vec{\Gamma}$ is a white noise term with the  dispersion of
$\sqrt{2\gamma k_BT}$, where $k_B$ is the Boltzmann constant and $T$ 
is the temperature. The dimensionless temperature, $k_BT/\epsilon$,
will be denoted by $\tilde{T}$.
The friction coefficient $\gamma$ is equal to $2m/\tau$ where  $\tau$
is a characteristic time scale. The dynamics are meant to be overdamped
so the characteristic time scale corresponds to a diffusional passage
of a molecular distance ($\sim$ 3 {\AA}) and is thus of order 1 ns. 
For small damping, $\tau$
would correspond to  time scale of (ballistic) oscillations in 
the Lennard-Jones well which is significantly shorter.
The equations of motion are solved by the fifth order 
predictor-corrector scheme \cite{Allen}.

As the average value of energy for hydrogen bond interaction in dsDNA 
we chose 0.6~kcal/mole, while in \cite{thiru1} it was chosen
around 0.5-0.7~kcal/mole, 
and in \cite{pablo}: 0.66~kcal/mole.
On average, 2.5 hydrogen bonds are created between the bases in dsDNA.
Hence the total average energy of interaction between paired bases in
the DNA helix is about 1.5~kcal/mole in our models.
This choice is consistent with $\tilde{T}=0.4$ corresponding to
$T$=300 K. The corresponding unit of the force, $\epsilon /${\AA} should 
be then of order 100 pN. 
In the entropic limit, the hydrogen bond potentials matter much less than
the thermal fluctuations. In our model, this starts to happen at 
$\tilde{T}$  of about 0.5 -- 0.6. 

\section{Model II: the 2-bead model description}

In the 2-bead model, we consider beads denoted by p and b
at each nucleotide as shown in Figure \ref{pbh23}. The p-beads are placed
in positions of the C4* atom and mimic the phosphate-ribose chain of the
DNA molecule. The harmonic tethering potential, as well as the bond and
the dihedral angle potentials are introduced in analogy to model I.
The b-bead in each nucleotide is placed in the geometrical
center of the base.
The absence of the h-beads of model I is compensated by introducing
hydrogen-bond-like interactions between the b-beads.
The average Cartesian coordinates of the model beads are provided in
Table~\ref{tabela}.
In order to distinguish between the A:T and G:C pairs in the DNA sequence, 
we strengthen the amplitude of the corresponding  Lennard-Jones potential
by the factor of 2 or 3 respectively.
The stacking potential, between the neighbouring b-beads along each DNA
is described as in model I.
The hard sphere diameters of the beads remain defined as in model I.
Average value of the bead mass is $\mathrm{m}= 162 $ g/mole and  
the distance between the beads that make effective hydrogen bonds is 5.5 $\pm$ 0.8. 
We have deteremined that the persistence length in model II at $\tilde{T}=0.4$
is about 50 nm.


\section{Model III: the 3-bead description}

Model III, introduced in ref. \cite{pablo} and shown schematically
in Figure~\ref{pbh23} provides a 3-bead description.
It differs from model II primarily as a result of a
different treatment of the backbone chain. In model II,
the sugar and phosphate groups of the same nucleotide are represented
by one bead, whereas in  model III the two groups are represented by 
separate beads so that the backbone chain is formed by connecting sugar 
bead (s) of one nucleotide to the phosphate bead of the
nucleotide that follows in the sequence.

The distinction between the phosphate and sugar groups is important in this 
model because it facilitates introduction of  electrostatic 
charges on the phosphate beads. The charges are introduced to describe 
interactions of the DNA with ions in the solvent but they also affect
the p-p distances through the resulting Coulombic repulsion.
The corresponding potential is given by
$V_{elec,ij} = \sum \frac{q_i q_j}{4\pi\epsilon _0 \epsilon r_{ij}}e^{-r_{ij}/\kappa _D}$,
where $\kappa _D$ is the Debye constant, and its value depends on the ionic strength
of the solution. For standard ionic strengths, $\kappa _D$ ranges from 11 to 15 {\AA}
(eg. when $\mathrm{[Na^{2+}]=50mM}$, one obtains $\mathrm{\kappa_D = 13.6}${\AA}). 
Here, we do not take this term into account, as its effect on the p-p
distances is minor and because our focus is on mechanical manipulations
and not on the effects resulting from variations of the ionic strength.

Other potentials used in this model are analogous to those used in models I and II.  
The exception
is the base pairing potential. We describe it by the effective Lennard-Jones
potential whereas Knotts et al.
\cite{pablo} by the 10-12 potential
\begin{equation}
V^{bp} = \sum_{base \; pairs} 4\epsilon _{bp} [5(\frac{\sigma_{bp}}{r_{ij}})^{12} - 
6(\frac{\sigma_{bp}}{r_{ij}})^{10}] \;\;,
\end{equation}
where $\epsilon_{bp}$ depends on the type of base pair (AT or GC, 
while $\epsilon_{GC}=3/2\cdot \epsilon_{GC}$), and $\sigma_{bp}$ is around 2.9 {\AA} for
all paired bases.


Hyeon and Thirumalai \cite{thiru1} have recently considered a model of 
the RNA hairpin in which every nucleotide is represented by three beads
which correspond to the  phosphate, sugar and base  groups  which is
analogous to model III of the double helix and to the model of
Knotts et al. \cite{pablo}.
Similar to ref. \cite{pablo}, the Debye-Hueckel potential between 
the phosphate beads is introduced to account for
screening by condensed counterions and for the hydratation effects.
However, there are differences pertaining to the
nonbonded potentials. In addition to the base pairing
potential, Hyeon and Thirumalai introduce a possibility
of stacking interactions between the base beads.
Such interactions do not arise between the successive
nucleotides, but may arise in, e.g., the head of an RNA hairpin.
These stacking interactions are responsible for existence
of the more complicated conformations, like the hairpin,
that the RNA may adopt.
As to the base pairing potential, Hyeon and Thirumalai
take the Lennard-Jones potential without making a
distinction between the number of the hydrogen bonds involved.
It affects the base beads which are within the distance of 7~{\AA}
and the corresponding depth of the potential is 1.8 kcal/mole.
Non-native base bead interactions are repulsive and
correspond to the energy parameter of 1.0 kcal/mole.

Another simple model of the DNA has been recently proposed by
Ouldridge et al. \cite{Doye} in the context of self-assembly
of DNA nanostructures in which the twisting character
of the individual strands is disregarded. In this model,
every nucleotide is represented by a softly repulsive
sphere of diameter $l$=6.3 {\AA} and by another smaller
sphere attached (nearly) rigidly to the center of the
repulsive sphere at a distance of 0.3$l$ away from it
and perpendicularly to the backbone.
The smaller sphere provides a center of attraction to
another small sphere and thus plays the role of a base
in the DNA strand.
Four  types of the base site are considered  and the
Lennard-Jones attraction links the complementary bases.
Additionally, the model incorporates a monomer-to-monomer 
bending energy to provide stiffness. In the ground state,
two strands run parallel to each other like in a $\beta$ sheet
in proteins.

Throughout the paper, we use the open square, solid square, and triangle symbols
to denote results corresponding to models I, II, and III respectively.

A convenient way to characterize the unravelling process
is by providing the distance at which a given contact
breaks down for the last time. A contact is said to be broken
if the corresponding distance exceeds 1.5 times the length
parameter $\sigma$ in the Lennard-Jones potential associated with
the contact. For models II and III, where there is only one connection
between paired bases (through the b-beads) the contacts are labelled by 
the nucleotide number, $l$, as counted from the side that is being pulled. 
In the case of model I, where each nucleotide can participate in
either 2 or 3 contacts,
the graphical representation uses the ordinates of
$l-0.1$ and $l+0.1$ for residues A and T whereas it involves
$l-0.2$, $l$ and $l+0.2$ for residues G and C.

\section{Unzipping at constant speed in scheme A}

\begin{figure}
\begin{center}
\includegraphics[width=1.0\textwidth,angle=0]{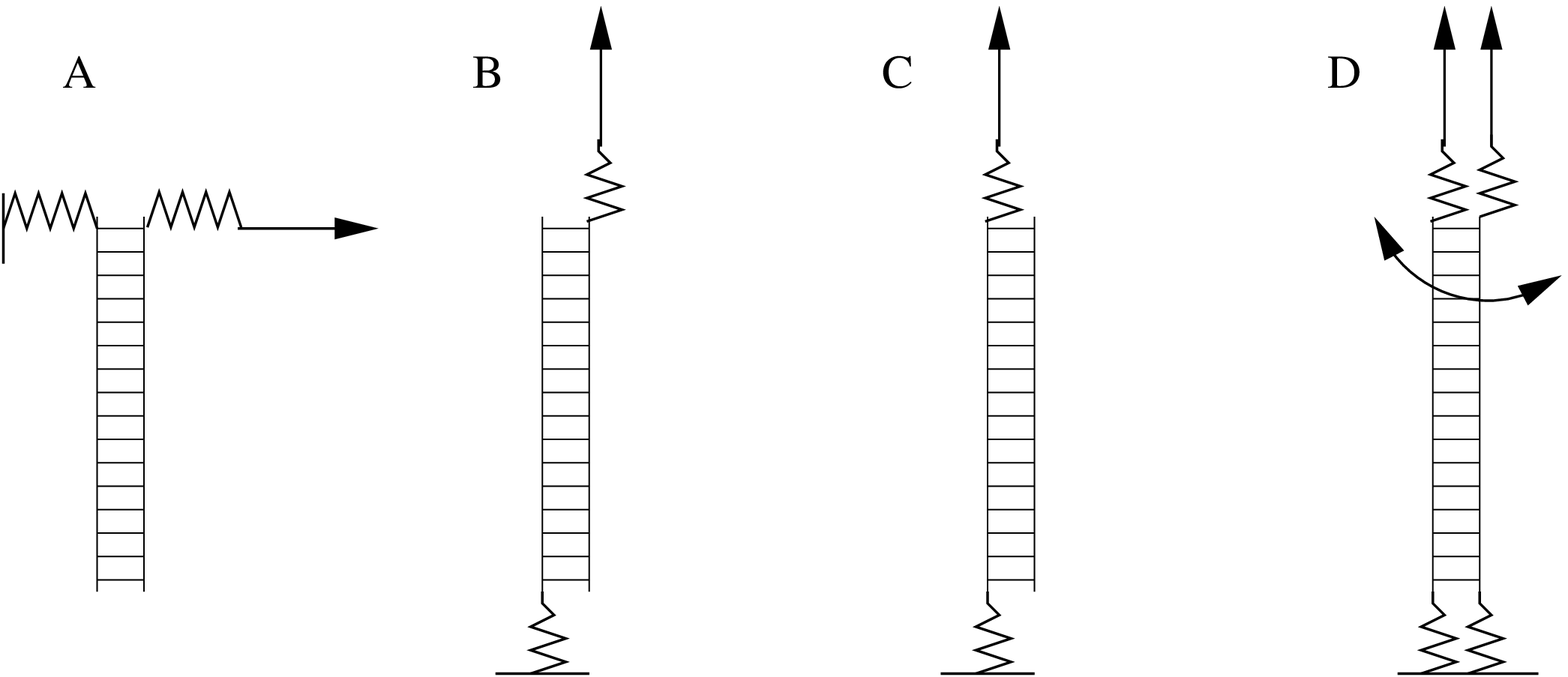}
\end{center}
\caption{Four possibilities of manipulation of the DNA double helix.}
\label{pbh}
\end{figure}

\begin{figure}
\begin{center}
\includegraphics[width=0.9\textwidth,angle=0]{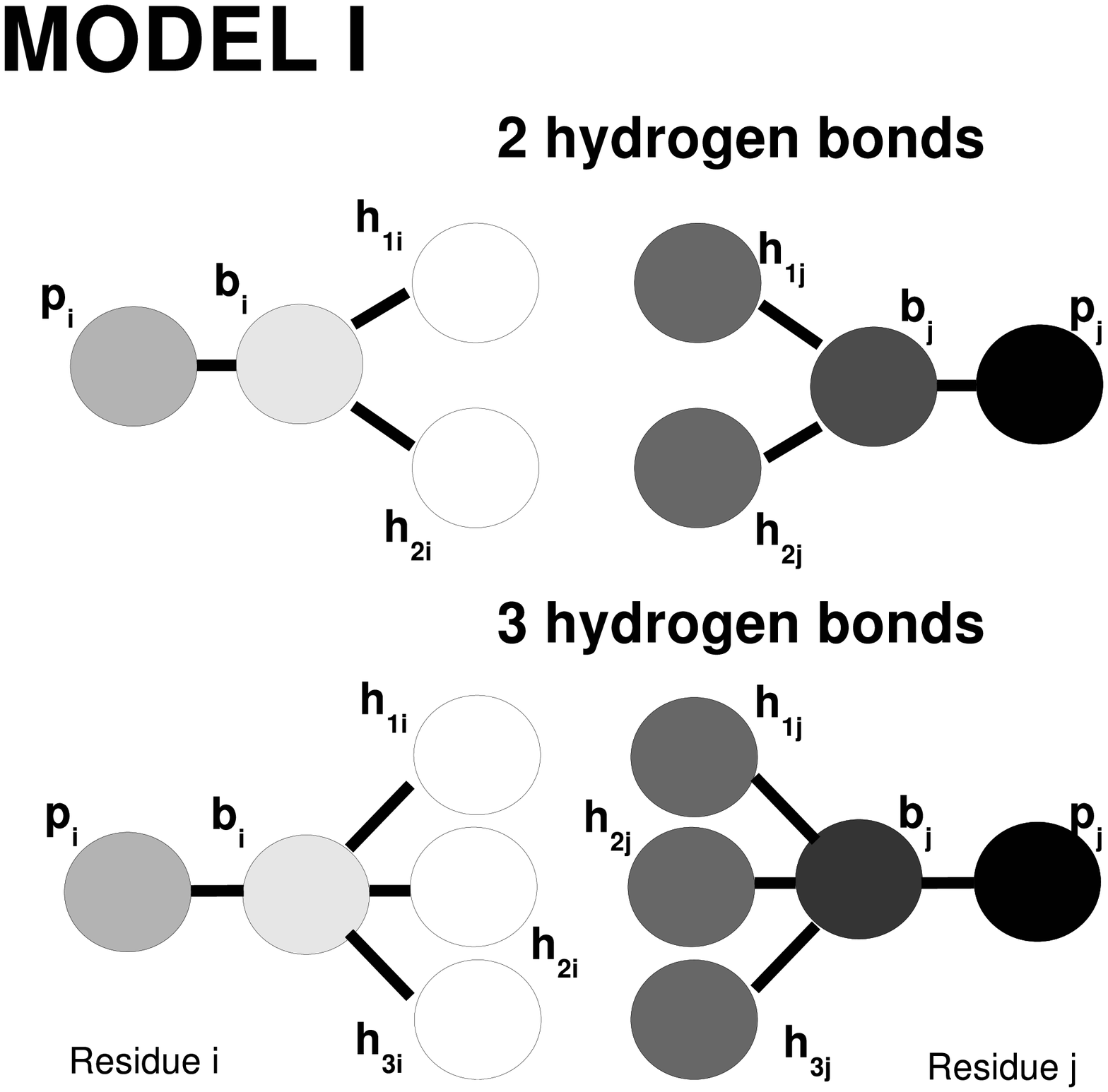}
\end{center}
\caption{A schematic representation of model I of  the dsDNA. 
         It shows formation of 2 and 3 hydrogen bonds.}
\label{pbh2}
\end{figure}

\begin{figure}
\begin{center}
\includegraphics[width=0.9\textwidth,angle=0]{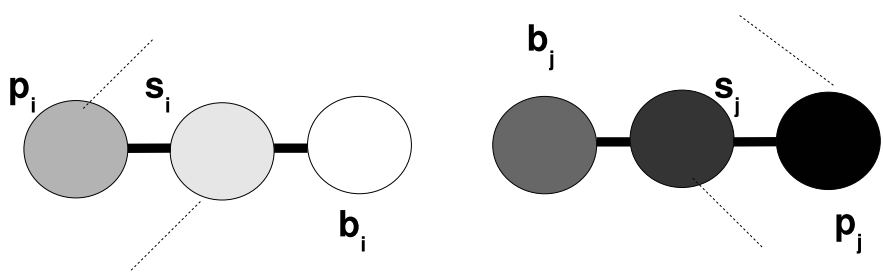}
\end{center}
\caption{A schematic representation of models II and III. The thin lines indicate
	the way the backbone chains are constructed.}
\label{pbh23}
\end{figure}

\begin{figure}
\begin{center}
\includegraphics[width=1.2\textwidth,angle=0]{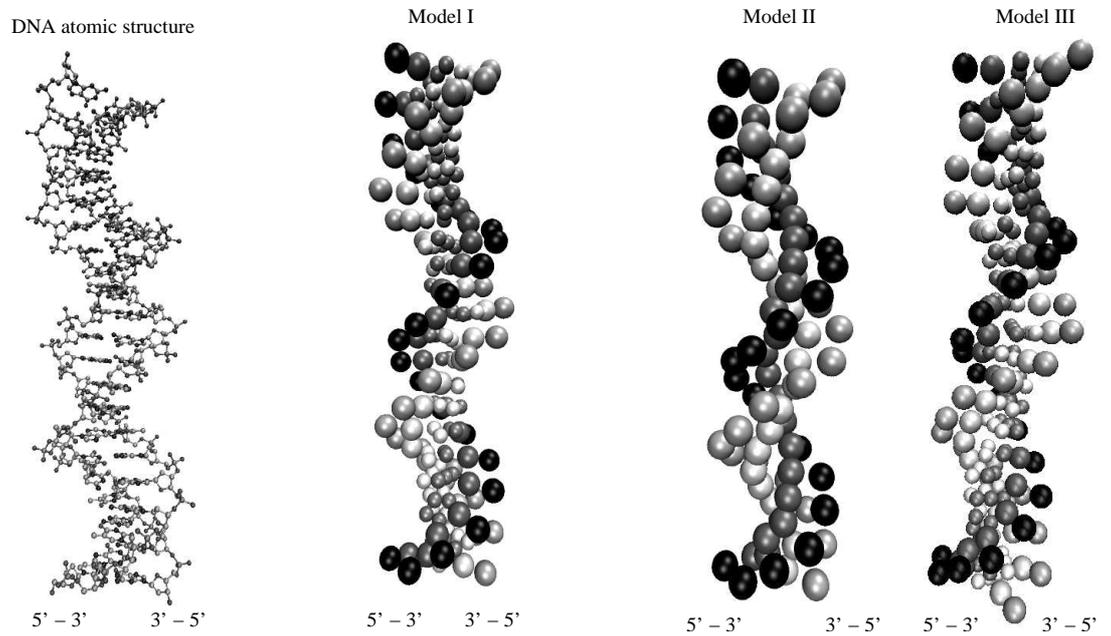}
\end{center}
\caption{The atomic representation of the 119D dsDNA structure is shown on the left.
         The remaining panels show the corresponding coarse grained  
         representations considered in this paper.}
\label{models}
\end{figure}

\begin{figure}
\begin{center}
\includegraphics[width=\textwidth,angle=0]{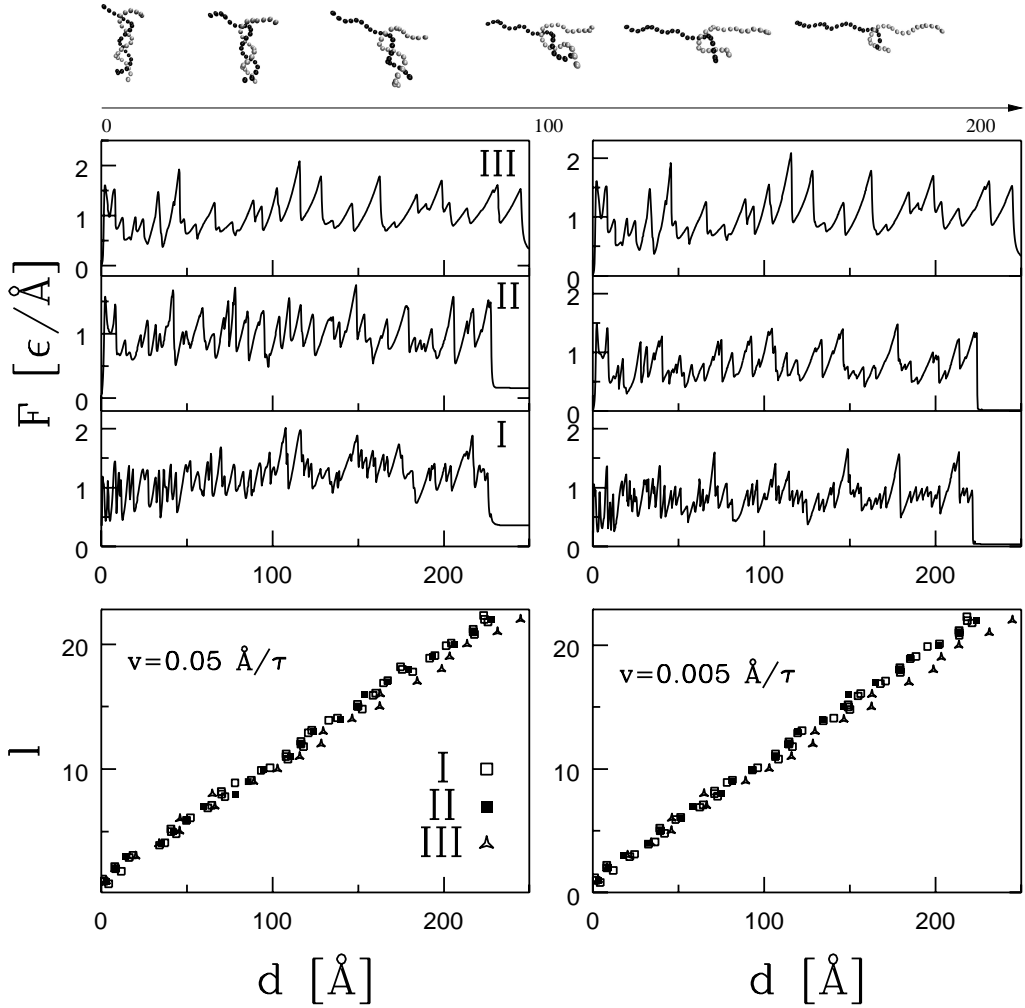}
\end{center}
\caption{The A-type stretching at  $T=0$. The snapshots at the top
show subsequent conformations of the pulled dsDNA in model II
for $v_p$=0.05 {\AA}/$\tau$. The panels below on the left correspond to
$v_p$=0.05 {\AA}/$\tau$ and those on the right to $v_p$=0.005 {\AA}/$\tau$
The middle panels show the $F-d$ curves for the three models as indicated.
At the end of the process, the two chains get fully separated and the
force drops to 0.
The bottom panels show the corresponding scenarios of unfolding.
}
\label{A0500}
\end{figure}

The stretching scheme A shown in Figure~\ref{pbh} leads to the 
unzipping process in which the hydrogen bonds
break by starting from the end that is being pulled. 
These bonds are enumerated by the index $l$ which is being counted
from the pulling end. Figure~\ref{A0500} 
shows the force vs. displacement patterns at
$T$=0 for two different values of the pulling velocity, $v_p$, of 0.05
and 0.005 {\AA}/$\tau$ and for the three models discussed. At
temperatures that are lower than $0.12~{\epsilon/k_B}$,  the $F-d$ 
patterns are qualitatively similar to the ones shown in Figure~\ref{A0500}
in the sense that the individual force peaks can usually be
related to unravelling of specific base pairs in the DNA sequence.

In the initial stages of unravelling,
all bonds that exist in the system get adjusted to some extent
and it is only later on that the unravelling process becomes
more site specific. This is well seen in model I for $d < 50$ {\AA},
especially for the higher pulling speed, where all force peaks
are quite similar.  In the other two models, this transient distance
is much shorter because the number of the adjustable bonds is smaller.
However, after this transient stage is passed, one can read off the pair 
sequence of the double helix from the $F-d$ patterns easily
because the higher peaks arise due to breaking of the G:C bonds.
The smaller the pulling speed the crisper the recognition of the sequence.

\begin{figure}
\begin{center}
\includegraphics[width=\textwidth,angle=0]{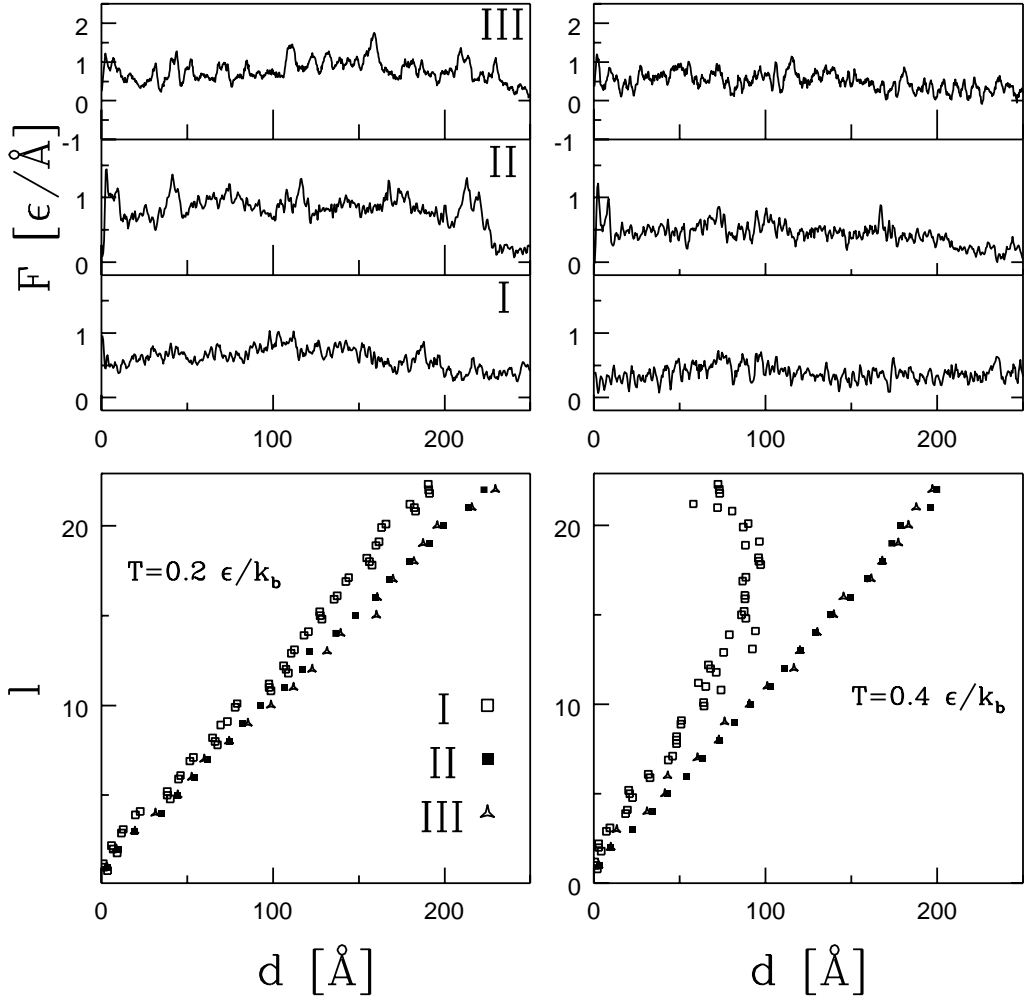}
\end{center}
\caption{Stretching in the A-type mode at $T=0.2 \epsilon /k_B$ (panels on the left) and
at $T=0.4 \epsilon /k_B$ (the panels on the right). The pulling velocity is 0.05 {\AA}/$\tau$.
}
\label{A0502}
\end{figure}

At $\tilde{T}=0.4$, 
we observe the peaks with the maximal value of 
0.39 $\epsilon/A$ (which corresponds to about 40 pN) - Figure~\ref{A0502}.
These values are about 20~pN higher than the 
experimental results \cite{unzip1:1997,unzip2:2002}, where for
$v_p = 200$~nm/s one obtains peaks of 18-20~pN 

The bottom panels of Figures \ref{A0500} and \ref{A0502} show
the scenario diagrams arising in scheme A.
At sufficiently low temperatures the zipping process is seen to be
proceeding linearly in time with minor differences in slopes
between the models. At higher temperatures, like for $T=0.4 \epsilon$/{\AA}
in model I, the scenario diagram data points acquire a curved appearance. 
This indicates that the thermal fluctuations rupture bonds at the idle 
(and not anchored) end of the
dsDNA before the mechanical unzipping process gets to them.

\begin{figure}
\begin{center}
\includegraphics[width=0.8\textwidth,angle=0]{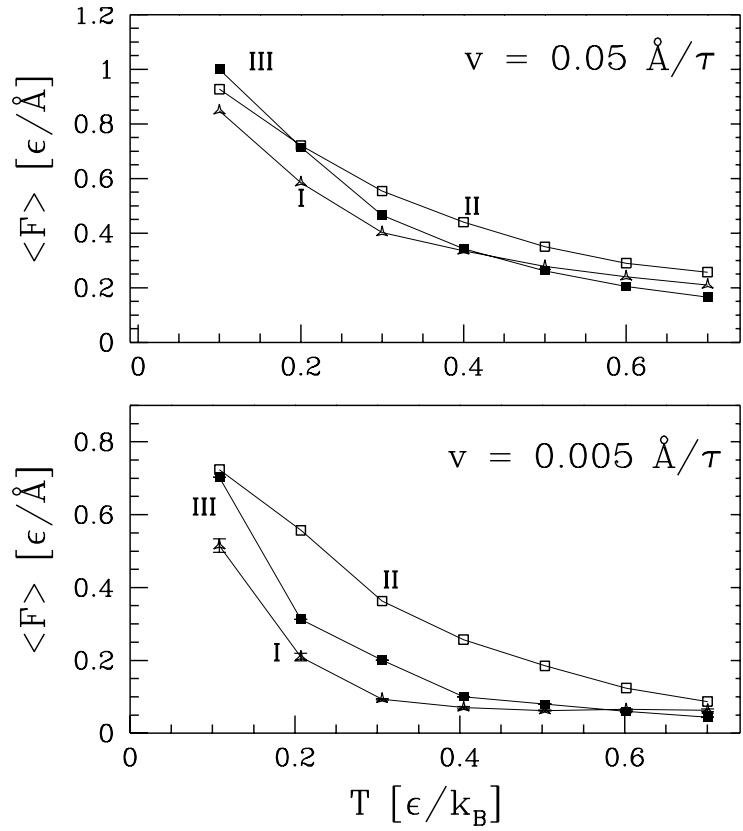}
\end{center}
\caption{The average force arising during the A-type stretching at 
$v_p$=0.05~{\AA}/$\tau$ and at $v_p$=0.005~{\AA}/$\tau$ as a function 
of $T$ for the three models.}
\label{Aavfor}
\end{figure}

For $T=0$ the force peaks have values of 1-2~$\epsilon$/{\AA} in all three
models. On increasing the temperature, the peaks and $<F>$, i.e. the force 
averaged over the duration of the full unravelling process (till $F$ drops to 0),
get lowered in a monotonic fashion, as shown in Figure~\ref{Aavfor},
which presents the results for unravelling with velocities
$v_p=0.05$~\AA/$\tau$ and $v_p=0.005$~\AA/$\tau$.
At the higher temperatures, model II yields the biggest mean forces,
independent of the pulling speeds.
Models I and III yield comparable mean forces at these
temperatures and it depends on the velocity which one is stronger
of the two. 

\begin{figure}
\begin{center}
\includegraphics[width=0.8\textwidth,angle=0]{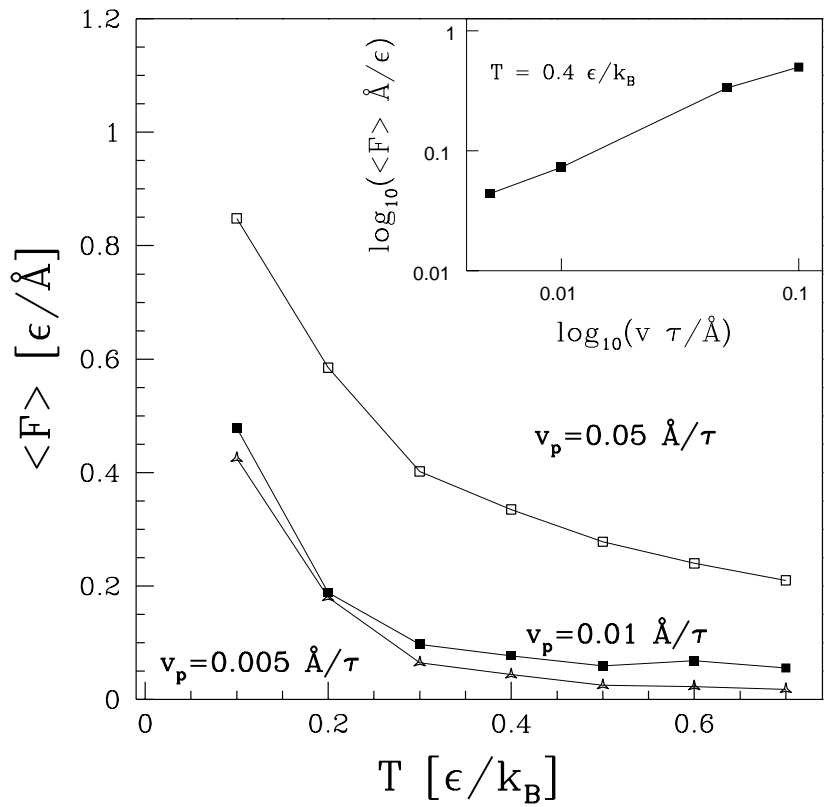}
\end{center}
\caption{The average force arising during the A-type stretching for three
different unravelling velocities in
model I. The inset shows the log-log plot of $<F>$ vs. $v_p$.}
\label{Aavfor2}
\end{figure}

With lowering the unravelling velocity, the calculated average forces 
decrease. In Figure~\ref{Aavfor2} we present the average forces for model I
for three different unravelling velocities. There is a big decrease of the $<F>$
between the $v_p=0.05$~\AA/$\tau$ and $v_p=0.01$~\AA/$\tau$, while the results 
obtained for $v_p=0.01$~\AA/$\tau$ and $v_p=0.005$~\AA/$\tau$ are close to 
each other. Generally, the smaller the velocity, the smaller the $<F>$ values. 
The similar dependencies are observed for models II and III.
Around $T=0.3$~$\epsilon$/{\AA} the rapid decrease in $<F>$
switches to a nearly constant behavior at higher temperatures.

\section{Stretching at constant speed in scheme B}

\begin{figure}
\begin{center}
\includegraphics[width=\textwidth,angle=0]{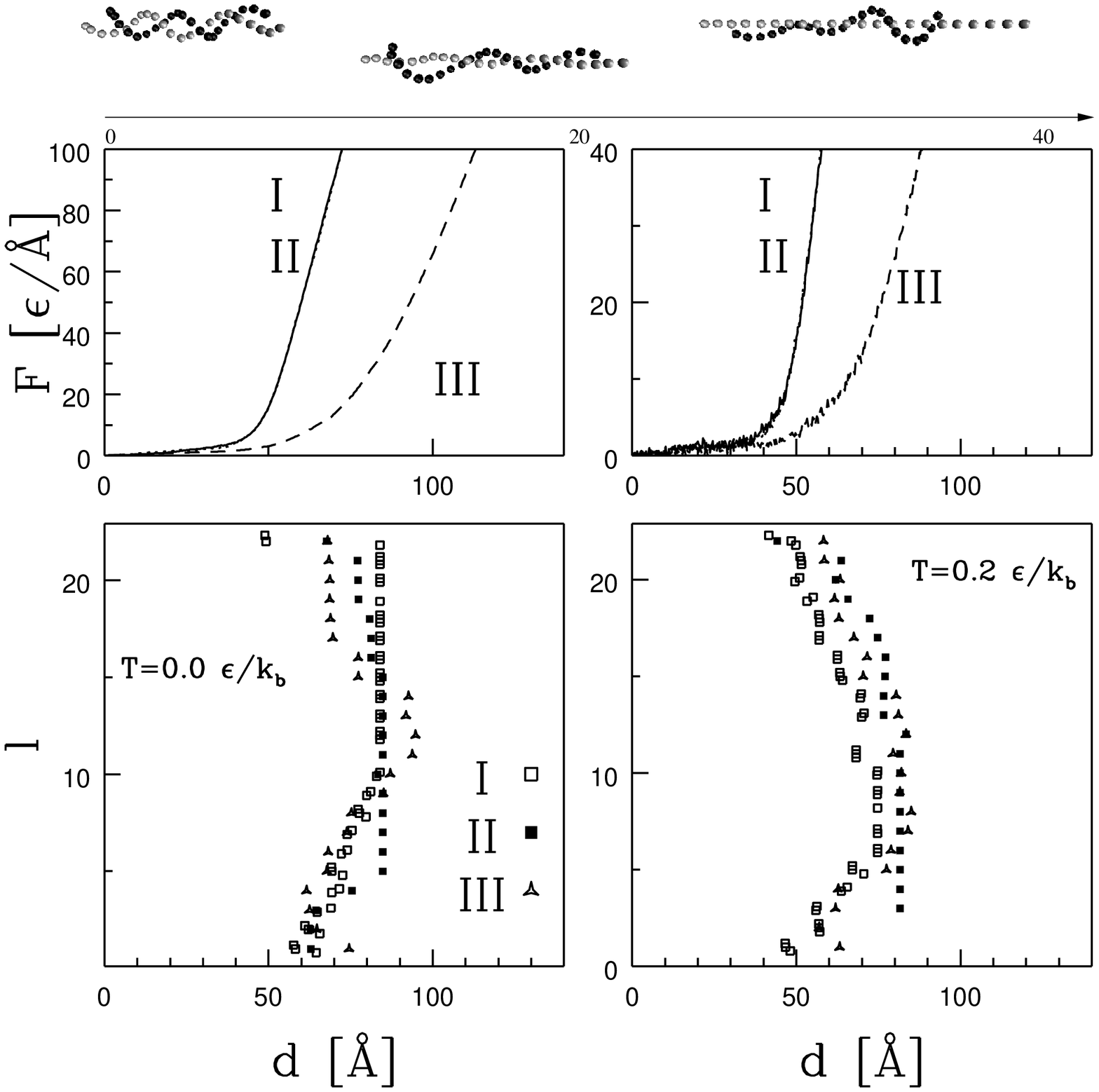}
\end{center}
\caption{The B-type stretching for $v_p$=0.05~{\AA}/$\tau$ at two different
temperatures for the three models. 
The panels on the left represent the results corresponding to
$T=0$, and the results shown on the right panels correspond to 
$T=0.2 \epsilon/k_B$. The snapshots presented at the top show conformations during the
stretching process of model II at $T=0.0$.  }
\label{B0502}
\end{figure}

The B-type stretching has an entirely different nature than 
the A-type one. In this scheme, only one chain undergoes
active stretching and this effect in turn influences the 
companion chain through the hydrogen-bond contacts.
Figure \ref{B0502} illustrates the mechanics of this
kind of manipulation. The snapshots (as obtained within model
II) show that a full extension of one chain results
in a substantial distortion of the other chain.
Furthermore, $F$  depends on $d$ in a monotonic fashion. There are
no force peaks even at $T$=0. The $F-d$ curves for models
I and II coincide and a bigger force arises faster than
in model III because of a more direct transmission of tension
between the p-beads.

The scenario diagrams also look distinct compared to scheme A
and in a way which is more sensitive to $T$.
At $T$=0 many contacts are broken nearly simultaneously.
At finite temperatures, the contacts at the extremities get
ruptured before unravelling of the contacts in the middle
in each of the models studied.
The higher the $T$, the earlier particular contacts break down.
We have observed insignificant dependence of the rupture
distances on the pulling velocity. The whole process 
results in unravelling both of the hydrogen-bond and
of the stacking contacts.

\section{Stretching at constant speed in scheme C}

\begin{figure}
\begin{center}
\includegraphics[width=\textwidth,angle=0]{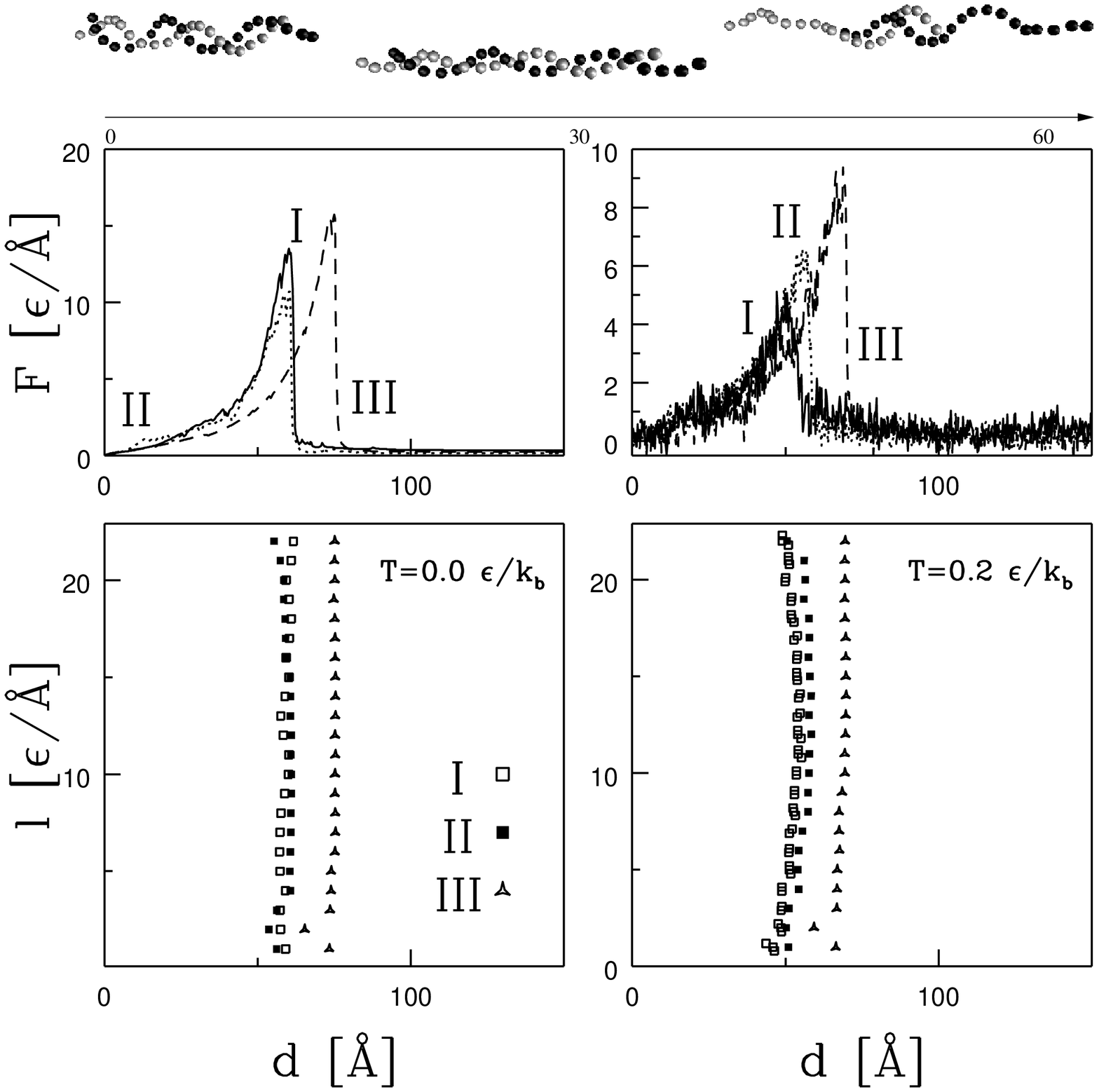}
\end{center}
\caption{The C-type stretching for $v_p$=0.05~{\AA}/$\tau$ at two different
temperatures for the three models. 
The panels on the left represent the results corresponding to
$T=0$, and the results shown on the right panels correspond to 
$T=0.2 \epsilon/k_B$. The snapshots presented at the top show conformations during the
stretching process of model II at $T=0$. }
\label{C0502}
\end{figure}

In the C-type stretching, one chain is made to slide along
its companion until the two chains separate as shown
in Figure \ref{C0502}. The $F-d$ curves display a major
peak which is an order of magnitude larger than the force
peaks observed in scheme A in all of the three models studied. 
The emergence of this major peak is due to an increasingly
cooperative resistance to manipulation of many contacts that are 
sheared simultaneously. Once the rupture takes place, the force drops 
down to the level corresponding merely to the thermal noise.
The cooperation level appears to be the greatest in model III,
followed by model II. In each of the models, the rupture
of all contacts is nearly simultaneous.

\begin{figure}
\begin{center}
\includegraphics[width=0.9\textwidth,angle=0]{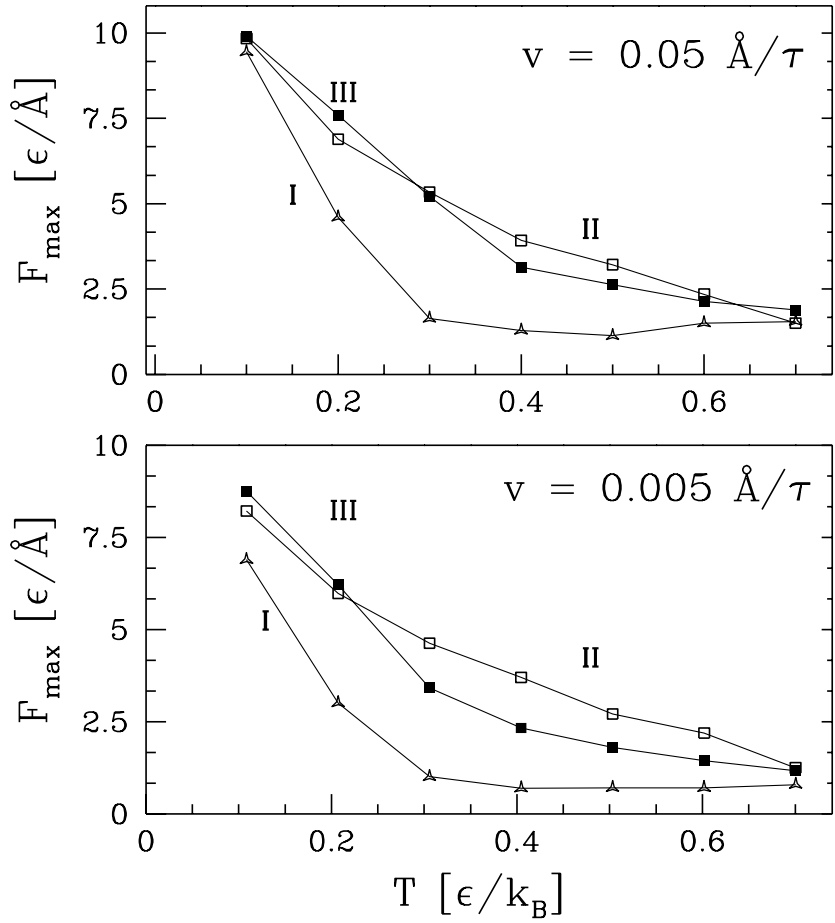}
\end{center}
\caption{The maximal force for the C-type stretching for $v_p$=0.05~{\AA}/$\tau$
and for $v_p$=0.005~{\AA}/$\tau$ as a function of 
temperature for the three models.}
\label{c1}
\end{figure}

\begin{figure}
\begin{center}
\includegraphics[width=0.9\textwidth,angle=0]{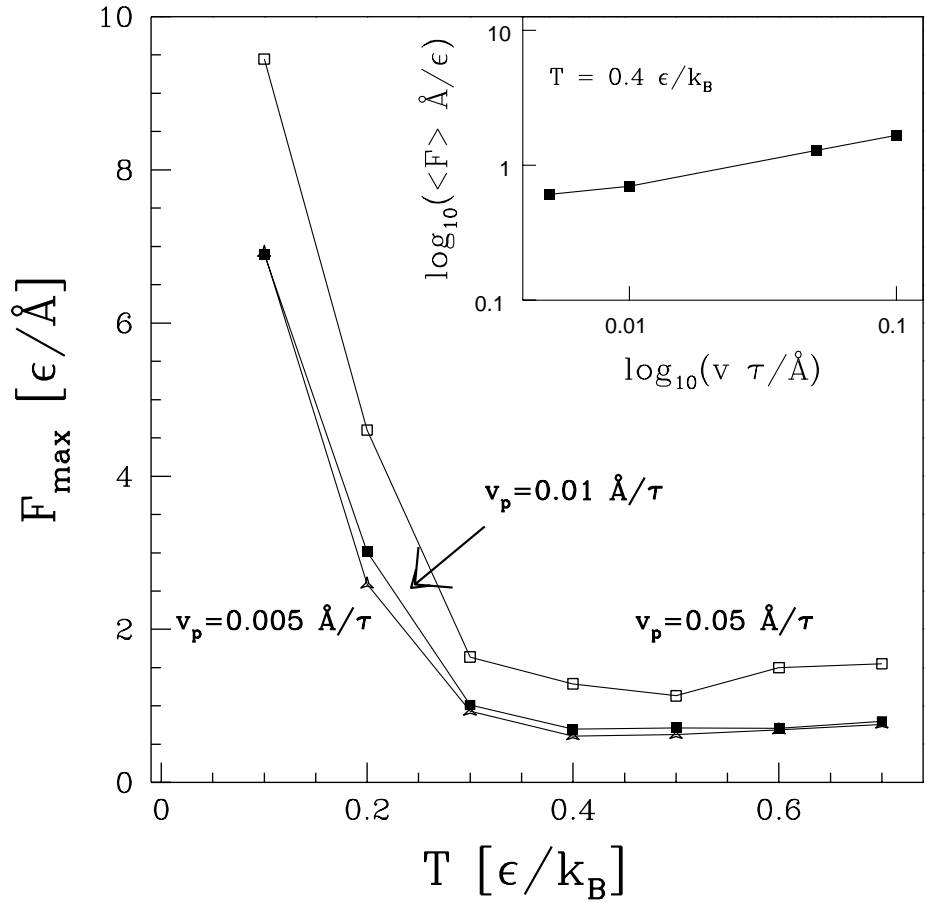}
\end{center}
\caption{The C-type stretching in model I. The main figure shows $F_{max}$
as a function of $T$ for three values of $v_p$.
The inset shows  the log-log plot of $F_{max}$ vs. $v_p$.}
\label{c3}
\end{figure}

The maximal force peak dependence on temperature shown in Figure~\ref{c1}
for two velocities indicated
is similar to what was observed for $<F>$ in scheme A, especially at low temperatures.
Models II and III yields are found to yield comparable forces which
are also noticeably larger than in model I except at the low
temperature end.
For the C-type stretching, the dependence on the stretching
velocity is weak as demonstrated in Figure~\ref{c3}.
The inset demonstrates that the dependence is nearly logarithmic.


\section{Stretching at constant force and constant torque in scheme D}

We now consider the tensile and torsional manipulations of the dsDNA
and focus on the determination of the corresponding force-torque 
phase diagram.

\begin{figure}
\begin{center}
\includegraphics[width=0.9\textwidth,angle=0]{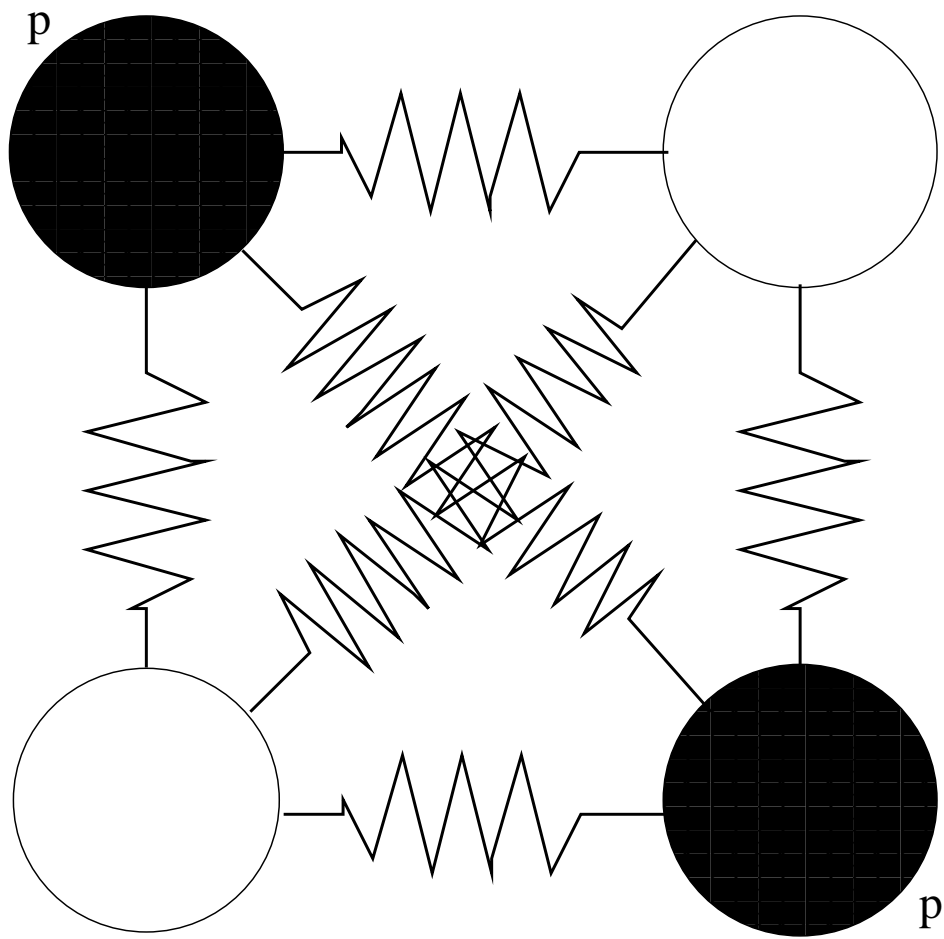}
\end{center}
\caption{A schematic view of the beads mounted at the DNA end in order
to apply torque through them. All beads are connected with the springs.}
\label{plane}
\end{figure}

We introduce the torsional stress of the dsDNA molecule in the 
following way. At one end of the molecule we choose two vectors
defining the plane. First vector is defined by the positions of
the extreme p-beads at the chosen end of the dsDNA. The second
vector defining the plane is a cross product of the first vector
and the dsDNA axis (which is defined by the midpoints of the
extreme beads on both ends of DNA molecule).
In the plane constructed in the way described above we add two more
beads as shown in Figure~\ref{plane} so that a square frame of four 
beads is formed. All beads in this frame are connected by the springs 
as in other structural bonds.
The extreme beads on the other end of the molecule are anchored at
their starting positions.

The torsion is applied to the DNA molecule by application of a force to
each of the four beads in the square frame. The torque is perpendicular
to the frame. The torsion is considered positive if it agrees with the
sense of the twist in the double helix in the B form and it is
considered negative otherwise.
The stretching force $F$ is a resultant force applied 
to all of the four beads in the frame and along the dsDNA axis.

\begin{figure}
\begin{center}
\includegraphics[width=0.9\textwidth,angle=0]{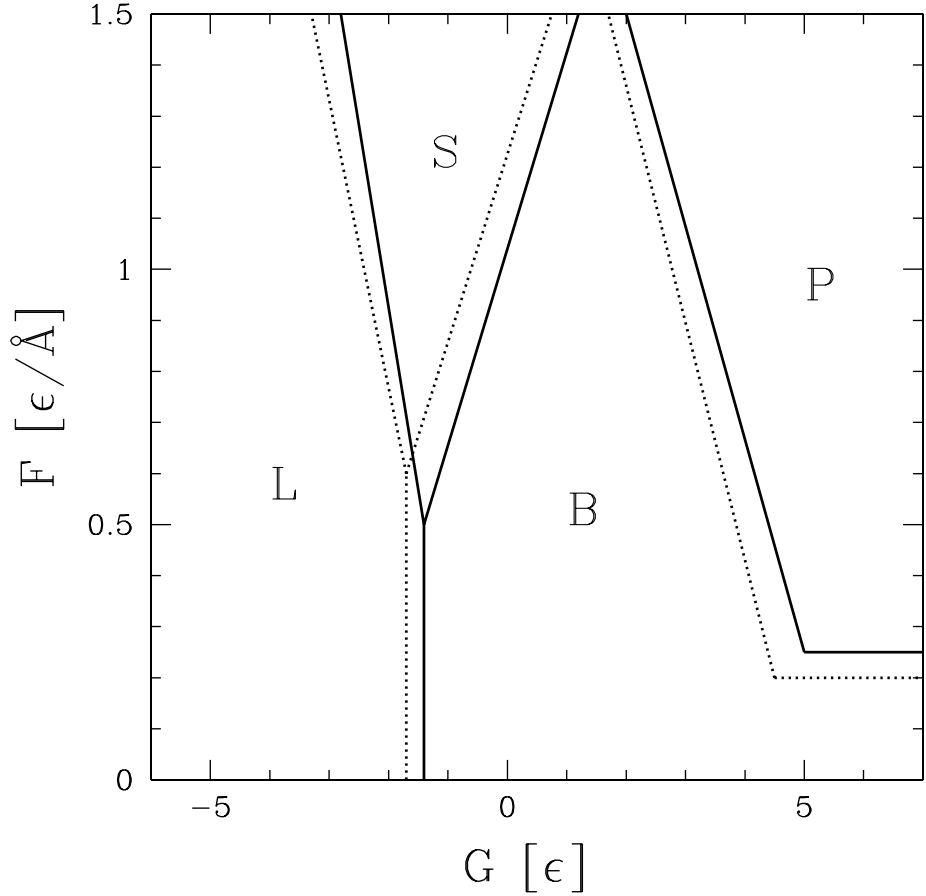}
\end{center}
\caption{
The phase diagram representing the final type of dsDNA structure obtained after
applying the stretching force, $F$, and the torqe, $G$, in model II. 
L denotes the L-DNA form, and
S signifies stretched chains, in which the backbones form straight lines. 
B denotes the original B-type structure. 
P corresponds to the Pauling form of the DNA, 
where the backbones get closer to each other, and bases stay outside 
of the helix. The solid lines represent results 
obtained for $\tilde{T}=0.4$, while the dashed ones for $\tilde{T}=0.2$.}
\label{m2fg}
\end{figure}

We first consider model II at two temperatures: 
0.2 and 0.4~$\epsilon$/{\AA}. The results are presented in the phase 
diagram in the Figure~\ref{m2fg}. The boundaries of the phases 
are approximate and are quite similar in both models.
These phase diagrams are also similar in appearance
to those established experimentally
\cite{tors1:2003,tors2:2006,tors3:2007} and theoretically
\cite{tote1,tors1:2003}.

We start from the dsDNA B-form
structure and observe the transitions into other phases of the
DNA structure. At T = 0.4~$\epsilon$/{\AA}, the DNA transforms into
the L-form under the torque $G$ of around $-1.5$~$\epsilon$.
At the lower temperature, this transformation
occurs for $G$ of around 1.8-2.0~$\epsilon$.
The simulations which lead to the L-form with the
value of the torque being close to 
this limiting value may last for up to $15000 \tau$. 
This time becomes significantly shorter for larger values of $G$.

The S-form of the DNA was experimentally characterised 
by overstretching of the dsDNA molecule by about $60\%$. 
In models I and II the elongation of the system
leads to  tightening of the bonds between p-beads along the chains, 
which finally leads to significant increase in the applied force. 
Thus the S-form region in the dsDNA phase diagram corresponds to structures
in which the backbone forms the straight line, without imposing the condition 
of $60\%$ overstretching. In T = 0.4~$\epsilon$/{\AA} such structures
occur while there is applied the force of  0.5~$\epsilon$/{\AA}. For
lower temperature there is needed a bit larger force for 
about 0.05-0.1~$\epsilon$/{\AA}. The above values were obtained for
applied torque $G$ of 1.5~$\epsilon$ and 1.7~$\epsilon$ respectively for
temperatures of 0.4~$\epsilon$ and 0.2~$\epsilon$.

The Pauling form is obtained when both the stretching force
and the positive twisting torque are large.
The smallest value of force needed to transform the system into 
the DNA P-form is $0.25~\epsilon$/{\AA}, while the torque
applied must be of value around $5.0~\epsilon$. For larger 
stretching forces, $G$ decreases to $2.1~\epsilon$ 
at $F=1.5~\epsilon$/{\AA}.
In the P-form form, the p-beads come closer together 
while the remaining beads (b and h) become exposed and face out of the helix.

\section{Stretching at constant angular speed in scheme D}

In order to study stretching at a constant angular speed,
we anchor the bottom beads and attach two frames to the top.
Each of these frames is as described in the previous section
and they coincide initially. The beads in one frame
are connected to their twins by elastic springs. As the outer
frame rotates at a constant angular speed, these springs
get stretched and impose a twist on the inner frame which is
glued to the DNA. This construct facilitates determination
of the resistive torque as it is accomplished by monitoring
stretching of the interframe springs.

\begin{figure}
\begin{center}
\includegraphics[width=0.9\textwidth,angle=0]{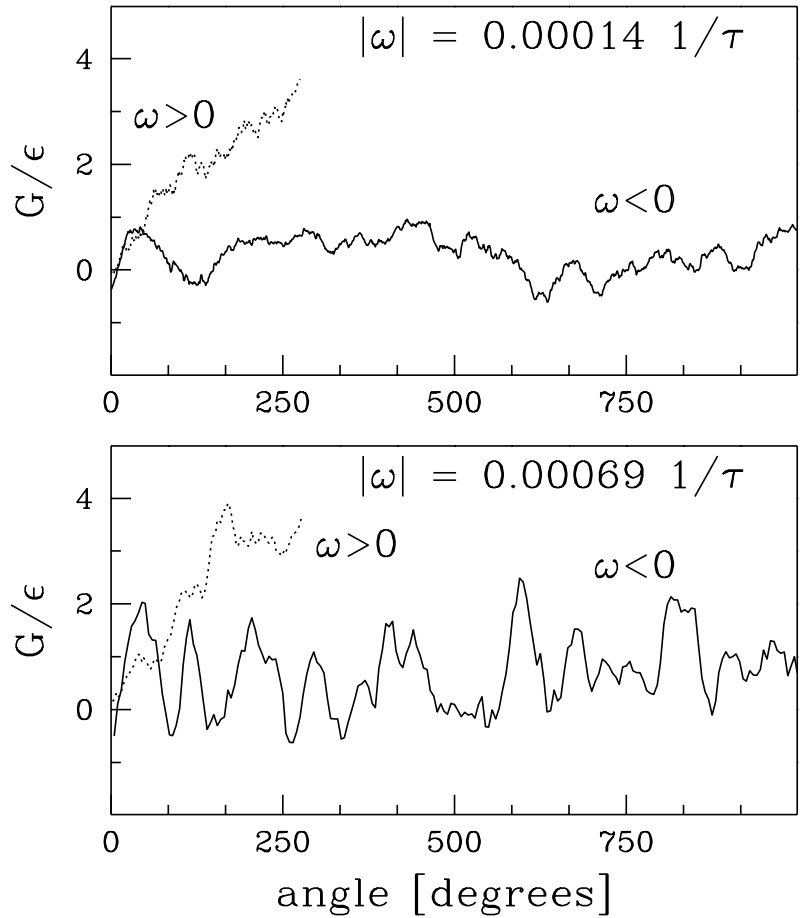}
\end{center}
\caption{
Torque of resistance to the twist as a function of the angle 
of rotation for two magnitudes of the angular velocities in $\tilde{T}=0.4$ as
indicated. The system consists of 20 base pairs and the simulations
have been performed within model II.
The dashed lines correspond to a sense of the twist that agrees with helical
rotation of the dsDNA. The solid lines correspond to the opposite sense of the twist.
 } 
\label{omega}
\end{figure}

Figure~\ref{omega} shows the torque of resistance to twisting
as a function function of the angle of rotation of the outer frame.
Two magnitudes of the angular speed, $\omega$, are used,
$0.00014\;\frac{1}{\tau}$ and $0.00069\;\frac{1}{\tau}$,
which differ by the factor of 5. 
We also probe two senses of the twisting: agreeing with the
helical twist ($\omega > 0$) or opposing it ($\omega < 0$).
The former leads to overtwisting and an indefinite growth
in the resistive torque due to an increasing infringement of the
steric constraints.
The latter results in unwinding and in a 
transition from the B-form to the L-form.
The results clearly depend on the twisting speed. In particular,
the average torques are $0.319\;\epsilon$ and
$0.890\;\epsilon$ for the smaller and faster negative angular
speeds respectively. 
The peaks in the torque result from the distortions,
but no contacts get ruptured when one infers about it
from the distance-based criterion.


%

%

\section{Conclusions}

The coarse-grained models of the dsDNA discussed here allow for studies of
features at the level of a single nucleotide. These models are found to be
fairly equivalent and indicate that sequence-specific events can be observed
in mechanical manipulations performed at low temperatures.
However, this capability becomes borderline around the
room temperature. 

Nevertheless the models proposed here should be useful when studying
DNA--protein complexes and when assisting
nanotechnological DNA assembly processes theoretically.
Examples of such processes 
are described in references \cite{future:2003,future:2004,future:2005}.

Many fruitfull discussions with P. Szymczak are appreciated.
This work has been supported by the grant N N202 0852 33 from the Ministry
of Science and Higher Education in Poland.

\newpage


\newpage 

\newpage

\newpage

\newpage 

\newpage



\end{document}